\documentstyle[aps]{revtex}

\input{epsf}

\begin{document}

\author{D. Blume}
\title{Fermionization of a bosonic gas
under highly-elongated confinement: A diffusion quantum
Monte Carlo study}

\date{\today }
\address{Department of Physics, Washington State University,
Pullman, Washington 99164-2814, USA
}

\maketitle

\begin{abstract}

The
diffusion quantum Monte Carlo technique is used to solve
the many-body Schr\"odinger equation
fully quantum
mechanically and nonperturbatively for
bosonic atomic gases in cigar-shaped confining potentials.
By varying the aspect ratio of the confining potential from 1
(spherical trap) to 10000 (highly elongated trap),
we characterize the transition from the
three-dimensional regime to the (quasi-)one-dimensional regime.
Our results confirm that the bosonic gas undergoes ``fermionization''
for large
aspect ratios.
Importantly, many-body correlations are included explicitly in our approach.
\end{abstract}

\draft


\section{Introduction}

\label{introduction}

One-dimensional systems
of impenetrable bosons are predicted to show intriguing behavior, namely
{\em{fermionization}}. An impenetrable one-dimensional gas
of $N$ bosonic atoms
behaves as if the system consisted of
$N$ one-dimensional
fictitious spin-less 
non-interacting fermions, a so-called Tonks or Tonks-Girardeau 
gas~\cite{tonk36,gira60,lieb63,lieb63a,olsh98,rojo99,gira00,petr00,kolo00,dunj01,gira01,gira01a,das02}.
Fermionization has profound effects on the characteristics of a system.
The ground state energy of
$N$ non-interacting bosonic
atoms in a one-dimensional harmonic trap with trapping frequency $\nu_z$ is
$E_0^{B,z}=N h \nu_z/2$. For comparison, the ground state energy of
$N$ non-interacting spin-less fermionic atoms under the same confinement is
$E_0^{F,z}=N^2 h \nu_z /2$, or $E_0^{F,z}=N E_0^{B,z}$.
In this paper we map out the transition of a bosonic
gas from three- to one-dimensional, and
monitor the signatures of
fermionization~\cite{petr00,dunj01,das02}.

A Tonks gas has not been observed experimentally yet,
and it is an open question how one would
probe a Tonks gas once it has been
created.
Effectively one-dimensional atomic gases may be realized
by confining $N$ atoms in highly elongated
cigar-shaped traps where the two tight confining
directions are characterized by the frequency
$\nu_{\rho}$,
and the
weak confining direction by $\nu_{z}$.
Correspondingly, the system exhibits two characteristic length scales,
the oscillator length in the $\rho$ direction,
$a_{\rho}=\sqrt{\hbar/(m \omega_{\rho})}$, and the oscillator
length in the $z$ direction, $a_{z}=\sqrt{\hbar/(m \omega_{z})}$,
where $\rho = \sqrt{x^2+y^2}$, $\omega_{\rho,z} = 2 \pi \nu_{\rho,z}$,
and $m$ the atomic mass of the trapped gas.
Such a confining potential has been realized experimentally by
G\"orlitz {\em{et al.}} with aspect ratios $L = \nu_{\rho}/\nu_{z}$
as large as 100~\cite{goer01} (see also~\cite{dett01,schr01,stre02}).
These experiments enter the quasi-one-dimensional regime,
which is characterized by $\hbar \omega_{\rho} \gg \hbar \omega_z$,
but not the truly one-dimensional regime.
Quasi-one-dimensional trapping confinement has also been realized
utilizing optical lattices~\cite{grei01} 
and wave guides~\cite{thyw99,bong01}.
Although experimental techniques will advance further,
there will always be an upper limit on the ratio $a_{\rho}/a_z$.
The natural question then is under what conditions does
the system behave truly one-dimensional, or,
to which extent is
the motion in the tight confinement directions decoupled from the motion
in the weak confinement direction.

Various authors have studied
the crossover from the three-dimensional regime to the
Tonks regime.
Common to these studies is some sort of variable separation
of the tight and loose confinement 
directions~\cite{dunj01,gira01,das02,tana00,mass02,meno02,sala02}.
For a cylindrical trap, e.g., Dunjko {\em{et al.}}~\cite{dunj01}
solve the Schr\"odinger equation
by employing the Lieb-Liniger model~\cite{lieb63}
and a local
density approximation.
For a toroidal trap, in contrast,
Das {\em{et al.}}~\cite{das02} 
develop a variational theory that factorizes the wave function
into a longitudinal and a transverse part.
Our approach here is different in spirit, in that it treats the
$3N$-dimensional Hamiltonian using the diffusion quantum Monte Carlo (DMC)
technique~\cite{ande75,reyn90a,hamm94,kosz96} 
fully quantum mechanically and nonperturbatively. Thus, no
separation of variables needs to be made, and
our results could be used to benchmark
variational calculations.

The DMC method has been previously applied to
microscopic Bose-Einstein condensates
(BECs) under spherical confinement~\cite{blum01}
(for applications of Monte Carlo techniques to
gaseous condensates 
see~\cite{krau96,grue97,pear98,gior99,dubo01}).
Similarly to Ref.~\cite{blum01}, we
monitor the ground state energy and structural properties
as a function of the number of particles. In addition, we vary
the aspect ratio over a wide range, i.e., $L=1,\cdots,10000$,
which allows signatures of fermionization to be investigated.
Importantly, our approach includes correlation effects explicitly
by treating all $3N$ coordinates of the system
equivalently.

Section~\ref{system} outlines the DMC formalism and
describes the system under study, while
Sec.~\ref{results} presents our results for the energetics
and structural properties.
Finally, Sec.~\ref{conclusions} concludes.

\section{Numerical techniques and characterization of the system}
\label{system}
\subsection{Diffusion quantum Monte Carlo technique}
\label{system_dmc}
The DMC method shows a favorable scaling with the number
$N$ of particles,
and allows bosonic many-body
systems to be treated fully quantum mechanically and nonperturbatively.
The DMC method~\cite{ande75,reyn90a,hamm94,kosz96}
solves the {\em{time-independent}} Schr\"odinger equation
\begin{eqnarray} \label{eq0a}
    H \varphi_n = E_n \varphi_n
\end{eqnarray}
for its ground state, $n=0$,
by considering the {\em{time-dependent}} Schr\"odinger equation.
Introduction of imaginary time
$\tau = it/ \hbar$ is used
to remove the explicit dependence on ``$i$'',
\begin{eqnarray} \label{eq1a}
    \frac{\partial \Psi(\vec{r},\tau) }{\partial \tau} =
    \underbrace{ \frac{\hbar^2}{2m} \sum_{j=1}^N
    \nabla_j^2 \Psi(\vec{r},\tau)}_{\mbox{diffusion}} -
    \underbrace{[V(\vec{r}) - E_{ref}] \Psi(\vec{r},\tau)}_{\mbox{source/sink}}.
\end{eqnarray}
Here, $\vec{r}=(\vec{r}_1,\cdots,\vec{r}_N)$ collectively
denotes the nuclear Cartesian coordinates of the atoms in the trap.
$V(\vec{r})$ is the potential energy surface of the
system, which is assumed to be known for now.
$E_{ref}$ shifts the absolute energy scale
without changing
the solution to the time-independent Schr\"odinger equation,
Eq.~(\ref{eq0a}).

A random walk technique is then used to calculate the steady state
solution [eigenvector $\varphi_0(\vec{r})$
with eigenvalue $E_0$]
by propagating to large $\tau$:
$\lim_{\tau \rightarrow \infty} \Psi(\vec{r},\tau)
\rightarrow \varphi_0(\vec{r})$.
Typically, about 1000 ``walkers'' are simulated simultaneously.
A walker is characterized by its position in Cartesian space and
a weight factor indicating its ``importance''.
To perform the propagation in imaginary time $\tau$,
the short-time approximation to the Green's function is introduced, and
the kinetic energy part (``diffusion'')
and the potential energy part (``source/sink'')
of the Hamiltonian are simulated separately at each
time step $\Delta \tau$.

The DMC algorithm outlined in the previous paragraphs is often
refered to as DMC {\em{without importance sampling}} or
DMC {\em{with a constant trial wave function}}.
It turns out that a trial wave function $\psi_T(\vec{r})$
that approximates the true ground state wave function $\varphi_0(\vec{r})$,
can be exploited fruitfully for increased accuracy.
Introduction of $\psi_T$ does not change the solution to
the time-independent Schr\"odinger equation [Eq.~(\ref{eq0a})];
however, it does give an additional term (``drift'') in
the time-dependent Schr\"odinger equation,
\begin{eqnarray} \label{eq1b}
    \frac{\partial (\Psi \psi_T)}{\partial \tau} =
    \underbrace{
    \frac{\hbar^2}{2m} \sum_{j =1}^N
    \nabla_j^2 \left(\Psi \psi_T \right)}_{\mbox{diffusion}} &-&
    \underbrace{\frac{\hbar^2 }{m}\sum_{j=1}^N \nabla_j
    \left(\Psi \psi_T \nabla \ln \psi_T \right) }_{\mbox{drift}}
    \nonumber \\
     &-& \underbrace{\left[\frac{-\hbar^2}{2m}
    \psi_T^{-1} \left( \sum_{j=1}^N \nabla^2_j \psi_T \right) +
    V(\vec{r}) - E_{ref}
    \right] (\Psi \psi_T)}_{\mbox{source/sink}}.
\end{eqnarray}
At each
time step $\Delta \tau$,
the new drift (or force) term introduces an additional update of the
Cartesian coordinates of the walkers.
Incorporation of a ``good'' trial wave function $\psi_T$, namely
one that resembles the true ground state wave function
$\varphi_0(\vec{r})$ as closely
as possible,
leads to a large reduction of the time step error of the
DMC calculation~\cite{mcdo81,bian88,chin90,umri93a}.
A ``bad'' trial wave function, in contrast,
can lead to biased
sampling of the configuration space, and can, in the worst case, lead to a
seemingly stable solution that differs from
the true, stationary solution of the time-independent
Schr\"odinger equation, Eq.~(\ref{eq0a}).
Incorporation of a trial wave function $\psi_T$ 
is required in the present application
to obtain converged DMC results.
The algorithm with a non-constant trial wave function
is typically refered to as DMC {\em{with importance sampling}}.

The DMC algorithm results in
an estimate of the ground state energy $E_0$, which
is accompanied by a systematic time step error and
additional statistical noise due to the stochastic nature of
the simulation process.
The time step error can be reduced by performing a series of
calculations for different time steps $\Delta \tau$, and then extrapolating
$E_0(\Delta \tau)$
to the zero time step limit $\Delta \tau = 0$. The statistical noise
of the energy expectation value can be reduced by
performing longer simulation runs.

The
calculation of structural ground state
properties requires a little more care than
the calculation of the
ground state energy because
Eq.~(\ref{eq1a})
interprets the wave function itself as the density
of the system, whereas Eq.~(\ref{eq1b})
interprets the product $\Psi \psi_T$ as the density of the system
(see also Sec.~\ref{system_app}).
To obtain structural properties, such as the
radial distribution, from the DMC simulation
with importance sampling, one therefore either uses
approximate mixed estimators or
employs essentially exact
extrapolation techniques such as descendant
weighting~\cite{kalo67,liu74,barn91}.
The latter is employed in the present application.

\subsection{Application to trapped atomic gases}
\label{system_app}
This section discusses the application of the outlined
DMC formalism with importance sampling to trapped bosonic gases.
The DMC method
``interprets'' the wave function of the system as its
density, and its application
is thus restricted to bosonic systems with
a positive and nodeless
wave function~\cite{comment1}.
What are the implications for treating
condensed atomic gases by the DMC method?
Ideally, one would like to simulate a
condensate using a sum of ``exact'' two-body
atom-atom
interaction potentials, e.g., the Krauss-Stevens Rb-Rb
potential~\cite{krau90} (and possibly
three-body terms), however, it is well known
that the ground state of such a many-body interaction potential
corresponds to a ``metallic snowflake-like'' state.
The gaseous
condensate state is not the ground state
of the system but a highly-excited metastable
state with a
finite lifetime.
To treat metastable condensates by the DMC method we 
artificially modify the two-body interaction potential
so that the gas-like BEC state becomes the ground state.

The full
potential surface of the $N$ atom system is composed of an internal
and an external piece.
We parameterize the interatomic many-body potential surface
by a sum of two-body potentials $V_2$,
$\sum_{j>k} V_2(r_{kj})$, where $r_{kj}$ denotes the distance
between atom $k$ and atom $j$, $r_{kj} = |\vec{r}_j - \vec{r}_k|$.
In addition, each atom feels the cigar-shaped external confining potential
$V_{trap}(\vec{r}_k)=\frac{1}{2}m (\omega_{\rho}^2 \rho_k^2  +
 \omega_{z}^2 z_k^2 ) $.
Our many-body Hamiltonian $H$ for $N$ mass $m$ atoms
then
becomes
\begin{eqnarray}
    H= -\frac{\hbar^2}{2m} \sum_{j=1}^N
    \nabla_j ^2 +
    \sum_{j>k,k=1}^N  V_2(r_{kj}) +
    \sum_{j=1}^N V_{trap}(\vec{r}_j),
\end{eqnarray}
where the Cartesian position vector $\vec{r}_j=(x_j,y_j,z_j)$
is measured with respect to the center of the trap.
Our calculations are performed for
a simple two-body hardcore potential,
$V_2(r_{kj})=\infty$ for $r_{kj} < a$, and $V_2(r_{kj})=0$ otherwise,
where $a$ denotes the two-body $s$-wave scattering 
length~\cite{blum01,krau96,grue97,pear98,gior99,dubo01,huan57,lee57a}.

Our trial wave function $\psi_T$ is chosen to have the simple 
Bijl-Jastrow-type form~\cite{bijl40,jast55},
\begin{eqnarray}
    \label{eq_trial_wave}
    \psi_T(\vec{r}_1,\cdots,\vec{r}_N)=
    \left( \prod _{j>k,k=1}^N \phi(r_{kj}) \right) \times
    \left( \prod_{l=1}^N \Phi(\vec{r}_l) \right),
\end{eqnarray}
which has the proper symmetry for a bosonic system,
namely, exchange of two particles leaves $\psi_T$ unchanged.
$\phi(r_{kj})$ is determined by the solution to the
radial two-body scattering equation for $V_2(r_{kj})$.
For the hardcore potential, one obtains
$\phi(r_{kj})=(1-a/r_{kj})$ for $r_{kj}>a$, and $\phi(r_{kj})=0$
otherwise. Thus, $\phi$ prevents
two atoms from ``crashing into each other''. In contrast,
$\Phi$ is the part of
$\psi_T$ that keeps the atoms inside the trap.
 $\Phi(\vec{r}_l)$
has two
variational parameters, $p_1$ and $p_2$,
\begin{eqnarray}
    \label{eq_trial_wave_ext}
    \Phi(\vec{r}_l) = \exp \left[-\frac{1}{2}
    \left(\rho_l / a_{\rho} \right)^2 -
    p_1 \left(z_l / a_z \right)^{p_2} \right].
\end{eqnarray}
We
determine the  best parameters $p_1$ and $p_2$ 
by minimizing
the energy expectation value
$ \langle \psi_T | H | \psi_T \rangle / \langle \psi_T | \psi_T \rangle$
through
variational quantum Monte Carlo (VMC)
calculations~\cite{hamm94,dubo01}.
For an ideal bosonic gas ($a=0$), $p_1=1/2$ and $p_2=2$, while
for $a>0$, the optimal values for $p_1$ and $p_2$ 
depend on $a$, $N$,
and the strength of the confining potential.

Our calculations are performed for $N=2,\cdots,20$ with
$m=m(^{87}\mbox{Rb})$, $\nu_z = 77.78$Hz, a two-body
$s$-wave scattering length $a=100$a.u., and varying $\nu_{\rho}$,
$\nu_{\rho}=1 \nu_z,\cdots,10000\nu_z$.
The DMC calculations become increasingly demanding with
increasing aspect ratio $L=\nu_{\rho}/\nu_z$, since
the strength of the tight confining
direction predominently determines the timestep $\Delta \tau$.
To ensure an acceptance rate of over 98\% in the
acceptance/rejection step of the DMC algorithm~\cite{hamm94},
$\Delta \tau$ has to be decreased with increasing $\nu_{\rho}$ as it scales
roughly with $a_{\rho}$.
Thus, to achieve the same
total propagation time for a system
with $L=1$ and for a system with $L=10000$,
we need to perform
about two
orders of magnitude more propagation steps for the system
with $L=10000$ than for the system with $L=1$. 

The number of propagation steps
also determines to first order
the statistical uncertainty of any quantity
calculated by the DMC technique.
Assume we wish to determine DMC energies with a
statistical uncertainty of $<0.2$$\hbar \omega_z$.
For a bosonic gas with $N=10$ atoms, e.g., this implies
an accuracy of about 1\% for $L=1$, and of about $4 \times 10^{-4}$\%
for $L=10000$. 
Achieving convergence of structural properties is more demanding
than achieving convergence of the ground state energy.
The descendant weighting technique~\cite{kalo67,liu74,barn91}
employed here requires a delay
between the ``original sampling'' and the ``sampling that is being used for
reweighting purposes''.
To obtain converged structural properties,
this delay
is as big as 10000 for some of our simulations.

\section{results}

\label{results}

This section presents our DMC results.
Section~\ref{results_energetics} summarizes
energetics, while Sec.~\ref{results_structural}
discusses structural properties.
Section~\ref{results_comparison} relates our results
to those by Petrov {\em{et al.}}~\cite{petr00}
and by Dunjko {\em{et al.}}~\cite{dunj01}.
\subsection{Energetics}
\label{results_energetics}
\begin{figure}[p]
\centerline{\epsfxsize=3.0in\epsfbox{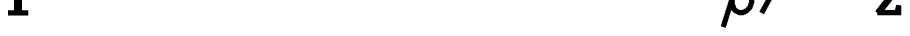}}
\caption{
\label{fig_energy_log}
Ground state DMC energy, $E_0^{DMC}$ (diamonds; for each $N$,
dotted lines connect the DMC energies to guide the eye), together
with the ground state energy obtained by solving the
two-dimensional Gross-Pitaevskii equation, $E_0^{GP}$ (triangles),
as a function of the aspect ratio $L = \nu_{\rho}/ \nu_z$ on a
log-log scale for $N=3,5,10$ and $20$. The statistical errorbar of
the DMC energies is smaller than the symbol size. }
\end{figure}

Figure~\ref{fig_energy_log} shows our DMC ground state
energies $E_0^{DMC}$ (diamonds; for each $N$, dotted lines connect the
DMC energies to guide the eye) in units of $\hbar \omega_z$
as a function of the aspect ratio $L$ on a log-log scale
for $N=3,5,10$ and $20$.
$L$ extends over four orders of magnitude,
and $E_0^{DMC}$ over about five orders of magnitude.
The statistical uncertainty of $E_0^{DMC}$ is smaller than the symbol size.
Figure~\ref{fig_energy_log} also shows the ground state
energies $E_0^{GP}$ (triangles)~\cite{comment2} obtained
by solving the mean-field Gross-Pitaevskii (GP)
equation~\cite{gros61,pita61,dalf98} in cylindrical coordinates
with interaction parameter
$\alpha = (N-1) a$,
\begin{eqnarray}
	\label{eq_gp}
   \left[
   -\frac{\hbar^2}{2 m} \nabla^2 +
   \frac{1}{2} m ( \omega_{\rho}^2 \rho^2 + \omega_z^2 z^2) +
   \frac{4 \pi \hbar^2 }{m}  \alpha | \chi_0(\rho,\varphi, z)|^2 \right]
   \chi_j(\rho,\varphi,z) = \epsilon_j \chi_j(\rho,\varphi,z).
\end{eqnarray}
The definition of $\alpha$ stems from the
derivation of the GP equation based on the Hartree-Fock 
formalism~\cite{esry97}.
$\epsilon_j$ and $\chi_j$ denote the orbital energy and the orbital wave 
function, respectively;
the GP ground state energy $E_0^{GP}$
can be obtained from the energy functional
$E_0^{GP}[\chi_0]$.
The numerical solution of Eq.~(\ref{eq_gp})
can be simplified by separating out
the motion in the
$\varphi$ coordinate.
On the scale of Fig.~\ref{fig_energy_log}, the DMC and GP energies are
indistinguishable.

\begin{figure}[tbp]
\centerline{\epsfxsize=3.0in\epsfbox{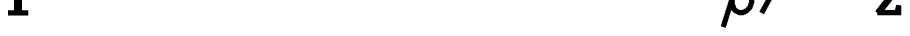}}
\caption{
\label{fig_energy_lin}
Energy per particle along the $z$ coordinate only. The DMC
energies $E_0^{DMC,z}/N$ (diamonds) are shown together with the GP
energies $E_0^{GP,z}/N$ (triangles) as a function of the aspect
ratio $L$ on a logarithmic scale for $N=3,5,10$ and $20$. To guide
the eye, dotted lines connect the DMC and GP energies 
for each $N$. Clear discrepancies between
$E_0^{DMC,z}/N$ and $E_0^{GP,z}/N$ are visible at larger $L$.
Horizontal solid lines on the right hand side show the energies in
the Tonks limit, $E_0^{F,z}/N$.}
\end{figure}

To interpret our results,
we subtract the ideal bosonic gas energy in the tight
confinement direction, 
$E_0^{B,\rho}=N h \nu_{\rho}$, from our DMC [GP] ground state energies,
and denote
the resulting energies
by $E_0^{DMC,z}$ [$E_0^{GP,z}$].
Figure~\ref{fig_energy_lin}
shows $E_0^{DMC,z}/N$ (diamonds)
as a function of $L$
on a
logarithmic scale.
To guide the eye, dotted lines connect
our data for each $N$.
The statistical uncertainty of the DMC energies $E_0^{DMC,z}$ increases with
increasing aspect ratio $L$.
E.g., the ground state energy of a gas with $N=5$
and
$L=10$
is $E_0^{DMC}=52.838(2) h \nu_z$.
Subtracting the ideal gas energy in the $\rho$ direction,
$E_0^{B,\rho}=50h \nu_z$,
gives a value of $E_0^{DMC,z}=2.838(2)h \nu_z$ with an uncertainty
of $0.002h \nu_z$.
For $L=10000$, we find
$E_0^{DMC,z}=12.4(2)h \nu_z$, a $100$ times larger statistical uncertainty.
Reducing the statistical uncertainty for $L=10000$ from
$0.2 h \nu_z$ to
$0.002 h \nu_z$ is possible in principle,
however, would require
significantly
more computer time.

For comparison, Fig.~\ref{fig_energy_lin} also shows
the GP energies $E_0^{GP,z}/N$ in the $z$ coordinate
only (triangles; for each $N$, dotted lines
connect these points to guide the eye). For small $L$,
the many-body DMC energies $E_0^{DMC,z}/N$
and the GP energies $E_0^{GP,z}/N$ agree well as expected.
However, as $L$ increases, significant deviations
occur. 
These deviations
are a consequence of considering the energy in the $z$ direction only
(see also Sec.~\ref{results_structural});
the deviations between the
total DMC and GP energies, $E_0^{DMC}$ and
$E_0^{GP}$, are negligibly
small (see Fig.~\ref{fig_energy_log}).
As we discuss now, Fig.~\ref{fig_energy_lin} confirms that
the systems under study
undergo fermionization for large $L$ as suggested by several
authors~\cite{petr00,dunj01,das02}.

Solid horizontal lines on the right hand side of
Fig.~\ref{fig_energy_lin} indicate the energy in the Tonks limit,
$E_0^{F,z}/N= N h \nu_z / 2$. For each $N$~\cite{comment3}, 
our DMC energies
$E_0^{DMC,z}/N$ approach the Tonks energy, suggesting that the
simulation parameters are such that the considered systems
behave {\em{fermionic}} in the $z$ direction, while the motion in the
$\rho$ direction is frozen out. Importantly, our simulations are performed
using a
trial wave function $\psi_T$, Eq.~(\ref{eq_trial_wave}), with the proper
bose symmetry for all $L$. 

\begin{figure}[tbp]
\centerline{\epsfxsize=3.0in\epsfbox{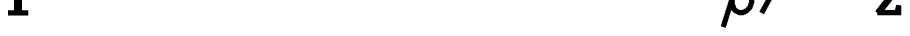}}
\caption{
\label{fig_energy_ratio}
Ratio between the DMC energy in the $z$ direction and the Tonks
energy, $E_0^{DMC,z}/E_0^{F,z}$ (lower part, diamonds) as a
function of $L$ on a logarithmic scale, together with the ratio
$E_0^{DMC,z}/E_0^{B,z}$ (upper part, triangles) for $N=2,3,5,10$ and
$20$. As $L$ increases the ratio $E_0^{DMC,z}/E_0^{F,z}$
approaches one, while the ratio $E_0^{DMC,z}/E_0^{B,z}$ approaches
values much greater than one, i.e., the value $N$,
indicating that the confined atomic
gas undergoes fermionization as $L$ increases.}
\end{figure}

To illustrate the fermionization transition further, the
lower part of Fig.~\ref{fig_energy_ratio} shows the DMC energies
$E_0^{DMC,z}$ divided by the Fermi energies $E_0^{F,z}$ as a
function of $L$ (diamonds) on a logarithmic scale
for $N=2,3,5,10$ and $20$. 
For each $N$, this ratio is close to
zero for small $L$, and approaches
one as $L$ increases. The $N=2$ system approaches the Tonks
limit at smaller $L$ values than the $N=20$ system. In
addition, the upper part of Fig.~\ref{fig_energy_ratio} shows the
ratio between the DMC energies $E_0^{DMC,z}$ 
and the ideal bosonic gas energy
in the $z$ direction, $E_0^{B,z}$ (triangles). For each $N$, this
ratio is close to one at small $L$, and increases with
increasing $L$.
In summary, the energy ratios shown in Fig.~\ref{fig_energy_ratio}
indicate that atomic gases under cigar-shaped confinement behave 
predominantly {\em{bosonic}}
for small $L$ and predominantly {\em{fermionic}} for large $L$.

Our analysis in this section depends profoundly on 
subtracting the ideal bosonic gas energy in the $\rho$ direction from the
many-body energy.
The next section discusses structural properties, for which
the behavior in the $\rho$ direction
and in the $z$ direction can be
{\em{naturally}} separated by
integrating over all coordinates but the coordinate of
interest, i.e., without making any approximations.

\subsection{Structural properties}
\label{results_structural} 
\begin{figure}[tbp]
\centerline{\epsfxsize=3.0in\epsfbox{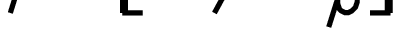}}
\caption{
\label{fig_density_rho}
DMC density $n_0^{DMC,\rho}$ as a function of $\rho$ for $N=5$ for
two different aspect ratios $L$, $L=10$ (diamonds) and $L=10000$
(triangles). To collapse the two DMC densities onto the same
scale, $\rho$ is measured in units of $a_{\rho}^{-1}$ with
$a_{\rho}=7307$a.u. for $L=10$, and $a_{\rho}=231$a.u. for
$L=10000$, respectively. Additionally, a solid line shows the
ideal bosonic gas density
[Eq.~(\protect\ref{eq_density_bose_rho})]. The DMC densities follow
the ideal bosonic gas density closely. }
\end{figure}

Figure~\ref{fig_density_rho} shows the
DMC density $n_0^{DMC,\rho}$, normalized such that
$\int_0^{\infty} n_0^{DMC,\rho}(\rho)  d \rho = 1$,
for $N=5$ for two different aspect ratios, i.e., $L= 10$
(diamonds) and $L=10000$ (triangles). The $\rho$ coordinate is
measured in units of $a_{\rho}^{-1}$ with $a_{\rho}=7307$a.u. for
$L=10$, and $a_{\rho}=231$a.u. for $L=10000$, respectively
(see Tab.~\ref{tab1}). The
statistical uncertainty of $n_0^{DMC,\rho}$ is largest for small
$\rho$ since the DMC walk samples the 
small $\rho$
region less than
the large $\rho$ region. A solid
line shows the density for the ideal bosonic gas, 
\begin{eqnarray} \label{eq_density_bose_rho}
    n_0^{B,\rho}(\rho) = \frac{2}{\sqrt{\pi} a_{\rho}}
    \exp \left[- (\rho / a_{\rho} )^2 \right].
\end{eqnarray}
On the scale of Fig.~\ref{fig_density_rho},
the DMC densities and the ideal gas
density are indistinguishable, which suggests that the motion along
the $\rho$ coordinate is indeed frozen out as $L$ approaches
large values (see Sec.~\ref{results_energetics}).

\begin{figure}[tbp]
\centerline{\epsfxsize=3.0in\epsfbox{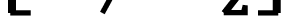}}
\caption{
\label{fig_density_z3}
DMC density $n_0^{DMC,z}$ for $N=3$ as a function of $z$ for four
different aspect ratios, $L=10$ (diamonds), $100$ (crosses),
$1000$ (triangles), and $10000$ (stars). $z$ is measured in units
of $a_z^{-1}$, where $a_z = 23107$a.u.. In addition, a grey solid
line shows the ideal gas density for bosons
[Eq.~(\protect\ref{eq_density_bose_z})], and a dark solid line
shows that for fermions [Eq.~(\protect\ref{eq_density_z})]. The
DMC density is close to $n_0^{B,z}$ for small $L$, and approaches
$n_0^{F,z}$ as $L$ increases. }
\end{figure}

The density along the $z$ coordinate is more interesting than the
density along the $\rho$ coordinate, since its behavior changes
dramatically as a function of the aspect ratio $L$.
Figure~\ref{fig_density_z3} shows the DMC density
$n_0^{DMC,z}(z)$, normalized such that $\int_{-\infty}^{\infty}
n_0^{DMC,z}(z) d z = 1$, for $N=3$ for four different aspect
ratios $L$, i.e., $L=10$ (diamonds), $100$ (crosses), $1000$
(triangles), and $10000$ (stars). 
$z$ is measured in units of
$a_z^{-1}$, where $a_z=23107$a.u..
For each $L$, dotted lines
connect the symbols to guide the eye. 
Additionally, a solid grey line shows the one-dimensional ideal
bosonic gas density $n_0^{B,z}$,
\begin{eqnarray} \label{eq_density_bose_z}
    n_0^{B,z}(z) = \frac{1}{\sqrt{\pi} a_z}
    \exp \left[- (z / a_{z} )^2 \right].
\end{eqnarray}
A solid dark line shows the ideal 
fermionic gas density $n_0^{F,z}$,
which is given by the sum of squares of the
single-particle wave functions,
\begin{eqnarray}
    \label{eq_density_z}
    n_0^{F,z}(z) = \frac{1}{N} \frac{1}{\sqrt{\pi} a_z} \sum_{k=0}^{N-1}
    \frac{1}{2^k k!} H_k^2(z/ a_z) \exp \left[-(z / a_z)^2 \right],
\end{eqnarray}
where the $H_k$ denote Hermite polynomials. The Fermi density
shows three distinct maxima, corresponding to $N=3$.

\begin{figure}[tbp]
\centerline{\epsfxsize=3.0in\epsfbox{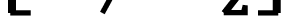}}
\caption{
\label{fig_density_z3_2}
DMC density $n_0^{DMC,z}$ for $N=3$ as a function of $z$ for
$L=10000$ (diamonds; the statistical
uncertainty is indicated exemplary for 
$z=1a_z^{-1}$). $z$ is measured in units of $a_z^{-1}$ with
$a_z = 23107$ a.u.. In addition, a dark solid line shows the fermionic
density $n_0^{F,z}(z)$ [Eq.~(\protect\ref{eq_density_z})], and a
grey solid line the square of the trial wave function component in the
$z$ direction [Eq.~(\protect\ref{density_trial})]. The dashed line
shows the density profile for
the Tonks limit [Eq.~(\protect\ref{density_dunjko})].}
\end{figure}

To ensure an unbiased
observation of the fermionization effect in
our DMC calculations, we choose a
``structureless'' trial wave function $\psi_T$,
that is a trial wave function without
oscillations. 
Figure~\ref{fig_density_z3_2} shows the $z$ component of $\psi_T$
squared (grey solid line),
\begin{eqnarray}
    \label{density_trial}
    M \exp \left[ -2p_1 (z / a_z) ^{p_2} \right],
\end{eqnarray}
with $p_1=0.018$ and $p_2=4.43$ ($M$ is a properly choosen
normalization constant), together with the DMC density for
$L=10000$ and $a=100$a.u. (diamonds) and the ideal fermionic 
gas density [Eq.~(\ref{eq_density_z}), black solid line].
We  checked the ``delay'' employed in the descendant
weighting scheme of our DMC calculation 
(see Sec.~\ref{system_app}), and estimate the
statistical uncertainty to be about three times the
symbol size.
The DMC density for $L=10000$ 
does not seem to show any 
Fermi-oscillations (note, however,
that the statistical error of the DMC density
is non-negligible, as indicated exemplary for one data point at
$z=1a_z^{-1}$).

Based on
a single-particle picture,
Petrov {\em{et al.}}~\cite{petr00} and 
Dunjko {\em{et al.}}~\cite{dunj01} 
derive an expression for the density of a gas 
under cylindrical confinement in the Tonks limit, 
\begin{eqnarray} \label{density_dunjko}
    n_0^{Tonks}(z) = \sqrt{\frac{2}{\pi^2 N a_z^2}} 
	\left( 1- \frac{z^2}{2 N a_z^2}      \right)^{1/2}.
\end{eqnarray}
Notably, $n_0^{Tonks}$ is dependent on $N$ and $a_z$ only,
i.e., it is independent of $a$.
Figure~\ref{fig_density_z3_2} indicates that
$n_0^{Tonks}$ (dashed line)
follows the Fermi density 
[Eq.~(\ref{eq_density_z}), dark solid line] closely, except for the
oscillations, which cannot be described within a single-particle
picture. The DMC density $n_0^{DMC,z}$ agrees well with 
$n_0^{Tonks}$. Section~\ref{results_comparison} presents a detailed
comparison between our work and that by 
Petrov {\em{et al.}}~\cite{petr00}
and by Dunjko {\em{et al.}}~\cite{dunj01}.

\begin{figure}[tbp]
\centerline{\epsfxsize=3.0in\epsfbox{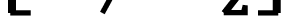}}
\caption{
\label{fig_density_z10}
DMC density $n_0^{DMC,z}$ for $N=10$ as a function of $z$ for four
different aspect ratios, $L=10$ (diamonds), $100$ (crosses),
and $1000$ (triangles). $z$ is measured in units
of $a_z^{-1}$, where $a_z = 23107$a.u.. In addition, a grey solid
line shows the ideal gas density for bosons
[Eq.~(\protect\ref{eq_density_bose_z})], and a dark solid line
shows that for fermions [Eq.~(\protect\ref{eq_density_z})]. The
DMC density is close to $n_0^{B,z}$ for small $L$, and approaches
$n_0^{F,z}$ as $L$ increases. }
\end{figure}

The Fermi density, Eq.~(\ref{eq_density_z}), changes significantly
with increasing $N$. To illustrate this effect,
Fig.~\ref{fig_density_z10} shows the DMC density 
$n_0^{DMC,z}$ as a function of $z$
for $N=10$ for three different $L$ values, $L=10$ (diamonds), $100$
(crosses), and $1000$ (diamonds),
using the same scale as in Fig.~\ref{fig_density_z3} for
$N=3$. The Fermi
density (dark solid line) now shows ten maxima, owing to the
increased number of terms in Eq.~(\ref{eq_density_z}). 
As $L$ increases the DMC density approaches the Fermi density. 
Our density studies, discussed
here in detail for $N=3$ and $N=10$, 
support
our
interpretation of the energetics presented in
Sec.~\ref{results_energetics}, and show
strong evidence that
an atomic gas under external confinement indeed undergoes fermionization
as $L$ increases.

\subsection{Comparisons with the literature}
\label{results_comparison}
This section relates our results to a set of criteria put
forward by Petrov {\em{et al.}}~\cite{petr00}
and by Dunjko {\em{et al.}}\cite{dunj01}.
Dunjko {\em{et al.}}\cite{dunj01}
derive five requirements
for a successful experimental observation
of the zero-temperature Tonks gas density profile,
Eq.~(\ref{density_dunjko}).
\begin{enumerate}
\item $a$ positive and $C1 = \frac{ C}{\sqrt{2}} \frac{a} {a_{\rho}} < 1$, 
	where $C \approx 1.4603$. 
	These conditions imply
	a positive one-dimensional coupling constant $g_{1d}$,
	\begin{eqnarray}
	g_{1d} = -\frac{2 \hbar^2}{m a_{1d}}
	\;\; \mbox{ with } \;\;
	a_{1d} = -\frac{a_{\rho}^2}{a} 
	\left( 1 - \frac{ C}{\sqrt{2}} \frac{a} {a_{\rho}} \right).
	\end{eqnarray}
\item $C2 = \frac{\hbar \omega_z N}{\hbar \omega_{\rho}} \ll 1$. 
	The transverse energy level spacing must exceed the energy 
	of the gas in the $z$ direction.
\item $C3 = \eta \ll 1$, 
	where $\eta$ denotes the governing parameter of the system,
	\begin{eqnarray}
		\eta = \frac{(9/2)^{1/3}}{8}
		\left[ \frac{4 a_{\rho}^4
		\left(1-\frac{C}{\sqrt{2}}\frac{ a}{a_{\rho}} \right)^2}
		{a^2 a_z^2} 
		\right]^{2/3} \; N^{2/3}.
	\end{eqnarray}
\item $C4 = \frac{\hbar \omega_z}{ 2 \hbar^2 / (ma_{1d}^2)} \ll 1$.
	The energy $\frac{2 \hbar^2}{m a_{1d}^2}$ must exceed the energy
	level spacing in the $z$ direction.
\item $N>1$. This condition is fulfilled trivially.
\end{enumerate}

Table~\ref{tab1} summarizes the characteristic lengths $a_{\rho}$
and the one-dimensional scattering lengths $a_{1d}$ as a function of
$L$ for $a=100$a.u., $m=m$($^{87}$Rb), and $\nu_z=77.78$Hz.
Condition~1. puts an upper limit on the aspect ratio $L$;
for the parameters chosen here, 
the coupling constant $g_{1d}$ becomes negative 
($a_{1d}$ becomes positive) for 
$L \approx 50000$.
$C1$ and $C4$ are independent of $N$, while
$C2$ and $C3$ scale with $N$ and $N^{2/3}$, respectively.

\begin{figure}[tbp]
\centerline{\epsfxsize=3.0in\epsfbox{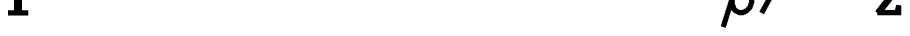}}
\caption{
\label{fig_criteria}
Criteria $C1$ (grey solid line), 
$C2$ (grey dotted lines), $C3$ (dark dashed lines), 
and $C4$ (dark solid line) 
as a function of the aspect ratio $L$
on a log-log scale 
using $a=100$a.u., $m=m(^{87}$Rb) and
$\nu_z=77.78$Hz (see Sec.~\ref{results_comparison}).
$C1$ and $C4$ are independent of $N$, while $C2$ and 
$C3$ scale with $N$ and $N^{2/3}$, respectively. 
$C2$ and $C3$ are shown for $N=3$ and $N=20$.}
\end{figure}

Figure~\ref{fig_criteria} 
shows the criteria $C1$ through $C4$ as a function of $L$ on a log-log
scale.
$C2$ and $C3$ are shown for $N=3$, and $20$.
$C2-C4$ are smaller than $0.1$ for about
$L  > 3500$, and smaller than $0.01$ for the extreme 
aspect ratio of about $L > 20000$.
The behavior of 
$C2$ and $C3$ indicates that
the ``window'', in which an atomic gas behaves like a
Tonks gas,
becomes narrower as $N$ increases. It thus seems advantageous
to experimentally work with small particle samples.
Toward this end,
optical lattices~\cite{grei01,grei02}, which allow
preparation of several 1000
identical copies of microscopic atom samples,
seem most promising.
Additionally tuning the two-body $s$-wave scattering length 
to large positive values through the use of Feshbach 
resonances~\cite{cour98,inou98}
would move the transition to the Tonks regime to
smaller aspect ratios.
A critical question, which is beyond the scope of this paper,
is whether the density profile in the $z$ direction
can be unambiguously measured in time-of-flight expansion
experiments, or whether mean-field effects may mask the signature of
the Tonks density profile~\cite{mass02,meno02,oehb02}.

\section{Summary and Outlook}
\label{conclusions}
The present DMC studies of atomic gases under external confinement 
with varying aspect ratio provide first
benchmark results, which
will allow the accuracy
of numerically less demanding, and more commonly used
approximate approaches to be tested. 
In contrast to variable
separation schemes and mean-field-type approaches,
the DMC formalism treats many-body correlations explicitly.
The DMC method consistently treats atomic gases under spherical
confinement and under highly-elongated cylindrical confinement,
and thus allows the transition from three-dimensional to
one-dimensional to be mapped out uniquely.
Application of the DMC technique can be extended to arbitrary
confining potentials such as toroidal or random potentials.
The main modification of the DMC algorithm 
would amount to adjusting
the $\Phi$ component of the trial wave function,
Eqs.~(\ref{eq_trial_wave}) and (\ref{eq_trial_wave_ext}), accordingly.

The DMC method in its present implementation
is restricted to treating many-body systems in their ground state.
To nevertheless treat metastable atomic gases, the present study
parameterizes the atom-atom interaction potential by a simple repulsive 
model potential, i.e., a hardcore potential. 
It would be interesting to investigate dependencies of the
energetics and structural properties on the detailed 
shape of the two-body interaction potential
as a function of the aspect ratio $L$.
Of particular interest would be the characterization
of atomic gases under external confinement as a function of the
ratios $a/a_z$ and $b/a_z$, where $b$ denotes the characteristic length
of the two-body potential~\cite{grib93,gao98,blum02}.

The
application of the zero-temperature
DMC technique in its present implementation
is restricted to a description of {\em{static properties}}:
We cannot make predictions regarding the lifetime or stability
of trapped atomic gases. Our studies here characterize 
atomic gases, assuming that they are stable.
Our approach provides no access to information related
to finite temperature effects,
time evolution, and coherence/decoherence phenomena.

In summary,
this paper presents exact many-body calculations for 
trapped atomic bosonic gases under external confinement with varying aspect
ratio. The trapping geometries considered range from spherical to 
highly elongated, $L=1-10000$.
The DMC technique is powerful in assessing the 
importance of many-body correlations.
Our study provides first benchmark results for
atomic gases under highly elongated cylindrical confinement.


\acknowledgments
Fruitful discussions with Chris H. Greene
at an early stage of this
project are gratefully acknowledged. I thank Brett D. Esry for
providing numerical solutions to the two-dimensional
Gross-Pitaevskii equation.

\twocolumn


\begin{table}[tbp]
\begin{tabular}{r|rrrr}
$L$ & $a_{\rho}$ [a.u.]  & $a_{1d}$ [a.u.] \\ \hline
     1. &      23107. & $-5.3 \times 10^6$ \\
     5. &      10328. &$-1.1\times 10^6$\\
    10. &       7307. & $-5.3 \times 10^5$     \\
    50. &       3268. & $-1.0 \times 10^5$     \\
   100. &       2311. &  $-5.1 \times 10^4$     \\
   500. &       1033. &   $-9.6  \times 10^3$    \\
  1000. &        731. &   $-4.6  \times 10^3$    \\
  5000. &        327. &   $ -7.3  \times 10^2$    \\
 10000. &        231. &    $-3.0  \times 10^2$    \\
 50000. &        103. &     $  -8.3 \times 10^{-2}$ \\
\end{tabular}
\caption{Characteristic length $a_{\rho}$ and one-dimensional scattering length
$a_{1d}$ for $a=100$a.u., $m=m(^{87}$Rb),
$\nu_z=77.78$Hz, and $a_z=23107$a.u. as a function of the aspect ratio $L$. }
\label{tab1}
\end{table}


\begin{thebibliography}{10}
\bibitem{comment1}
{Modifications of the DMC algorithm exist that allow treatment of excited
  states, however, their application to gaseous condensates is non-trivial.}
\bibitem{comment3}
{Obtaining convergence for systems with $N=20$ and large $L$ is numerically
  demanding. In this paper we only report results for $L \le 200$.}
\end{thebibliography}
\end{document}